# Magnetic-field sensitive charge density wave orders in the superconducting phase of UTe$_2$

Anuva Aishwarya[1], Julian May-Mann[1,2], Arjun Raghavan[1], Laimei Nie[1,2], Marisa Romanelli[1], Sheng Ran[3,4,5], Shanta R. Saha[3], Johnpierre Paglione[3], Nicholas P. Butch[3,4], Eduardo Fradkin[1,2], Vidya Madhavan[1]

[1]Department of Physics and Materials Research Laboratory, University of Illinois at Urbana-Champaign, Urbana, IL, USA.
[2]Institute for Condensed Matter Theory, University of Illinois, 1101 West Springfield Avenue, Urbana, Illinois 61801, USA
[3]Maryland Quantum Materials Center, Department of Physics, University of Maryland, College Park, MD, USA
[4]NIST Center for Neutron Research, National Institute of Standards and Technology, Gaithersburg, MD, USA.
[5]Department of Physics, Washington University in St. Louis, St. Louis, MO 63130, USA

**Abstract:**

**The intense interest in triplet superconductivity partly stems from theoretical predictions of exotic excitations such as non-abelian Majorana modes, chiral supercurrents, and half-quantum vortices. However, fundamentally new, and unexpected states may emerge when triplet superconductivity appears in a strongly correlated system. In this work we use scanning tunneling microscopy to reveal an unusual charge density wave (CDW) order in the heavy fermion triplet superconductor, UTe$_2$. Our high-resolution maps reveal a multi-component incommensurate CDW whose intensity get weaker with increasing field, eventually disappearing at the superconducting critical field, Hc$_2$. To explain the origin and phenomenology of this unusual CDW, we construct a Ginzburg-Landau theory for a uniform triplet superconductor coexisting with three triplet pair density wave (PDW) states. This theory gives rise to daughter CDWs which would be sensitive to magnetic field due to their origin in a triplet PDW state, and naturally explains our data. Our discovery of a CDW sensitive to magnetic fields and strongly intertwined with superconductivity, provides important new information for understanding the order parameter of UTe$_2$ and uncovers the possible existence of a new kind of triplet PDW order which has not been previously explored.**

In the ongoing search for new phases of matter, the heavy fermion superconductor Uranium ditelluride ($UTe_2$) which combines strong correlations and triplet-superconductivity (1-4) with possible non-trivial topology (5-7), is an extremely promising system. $UTe_2$ is paramagnetic, displaying no magnetic ordering down to the lowest temperatures (8), and superconducts below the critical temperature ($T_c$) of ~2 K (1,4). The unusually high upper critical field ($Hc_2$) (1), multiple field-reentrant superconducting phases (3), minimal change in the Knight shift (1,8), and exceptionally large Sommerfeld coefficient below $T_c$ (1), all provide strong evidence in support of unconventional triplet superconductivity. Recent measurements of a non-zero polar Kerr effect below $T_c$ (6) show that the superconducting state has broken time-reversal symmetry. Theoretical studies suggest that $UTe_2$ might be a topologically non-trivial Weyl superconductor (7,9,10) and harbor a chiral triplet state with Majorana arcs (11,12). This slew of intriguing phenomenology combined with indications of non-trivial topology, make $UTe_2$ an exciting and unique platform for the realization of fundamentally new states.

In this work, we use scanning tunneling microscopy/spectroscopy (STM/S) to study single crystals of $UTe_2$ below $T_c$. $UTe_2$ crystallizes into a body-centered orthorhombic structure with two Uranium atoms per unit cell (13). The unit cell consists of bi-trigonal prisms of U and Te where a U-U dimer is surrounded by two inequivalent Te atoms (based on U-Te bond lengths) labelled $Te_1$ and $Te_2$ in Fig. 1a (dark/light blue colors). The chains of bi-trigonal prisms run parallel to the a-direction and are offset by c/2 in the c-direction (Fig. 1b) where c is the height of the unit cell. The lattice may also be visualized as slabs of bi-trigonal prisms oriented along the (011) direction (Fig. 1b). The $UTe_2$ samples in this study were cleaved at temperatures of ~90 K and immediately inserted into the STM head (see supplement for details). Previous studies (5) have shown that (011) is the easy cleave plane and the atoms readily visible in the topography are the $Te_1$ and $Te_2$ atoms which appear as chains (Fig. 1c,d). A Fast Fourier transform (FFT) of the topography is shown in Fig. 1g where the Te-Bragg peaks/reciprocal lattice vectors are shown by the cyan dashed arrows. We denote the Te-Bragg peaks by, $q_{1,2}^{Te} = (\pm q_{o_x}, q_{o_y})$, where $q_{o_x}$ and $q_{o_y}$ denote coordinates in $x$ and $y$ directions in the FFT as labelled. Note that the $x$ and $y$ directions used here are for ease of notation and do not indicate the crystallographic directions. To complete the picture, we identify the (011) projection plane of the three-dimensional orthorhombic Brillouin Zone (BZ) in momentum space as shown in Fig. 1e. Constructing the BZ for the surface from the primitive lattice vectors gives rise to the elongated, hexagonal BZ as shown in Fig. 1g. The center of this BZ is labelled as S and the vertices are labelled as $L_1$, $L_2$ and W by convention (14).

As is evident from Fig. 1g, apart from the Te-Bragg peaks, there are three additional peaks near the $L_1$, $L_2$ and W points of the BZ in the FFT (enclosed by two squares and a triangle). To understand the origin of these extra peaks we perform spectroscopic imaging i.e., we obtain the differential tunneling conductance $\frac{dI}{dV}(r, E)$ maps as a function of energy ($E$). A $\frac{dI}{dV}(r, E)$ map is directly proportional the local density of states (LDOS) and can provide information about the band-structure through quasi particle interference as well as Fermi surface instabilities like CDWs. In addition to the atomic corrugation of the Te-lattice, our LDOS maps (Fig. 2a-c and Fig. S2) show modulations both above and below the Fermi Energy ($E_F$) which are distinct from the lattice. This additional modulation is also captured in the FFTs shown in Fig. 2d-f and gives rise to the same additional peak structure seen in the FFT of the topography shown in Fig. 1g.

To distinguish between signals from quasi particle interference and CDWs, we study the energy dependence of the q-vectors associated with the peaks in the FFT. To do this we obtain linecuts of the FFTs of the LDOS maps in the three important momentum space directions, S–$L_1$, S–$L_2$ and S–W, henceforth labelled as Line 1, 2 and 3 respectively, and plot this as a function of energy. This information is presented as an intensity map in Fig. 2g, h and i. Contrary to energy-dispersive features such as quasiparticle interference, we find that the magnitude of the three q-vectors shows no energy dependence. This indicates that the observed modulations arise from CDW order in this material. We have verified the existence of these CDWs across 11 different samples and tips from 3 different growth batches. We label the CDWs which are shown by the orange square, red square and purple triangle as $q_i^{CDW}$, where $i = 1, 2, 3$, respectively. The CDW q-vectors are: $q_1^{CDW} = (-q_{o_x}, 0.43 q_{o_y})$, $q_2^{CDW} = (q_{o_x}, 0.43 q_{o_y})$, $q_3^{CDW} = (0, 0.57 q_{o_y})$, where $q_{o_x}$ and $q_{o_y}$ are the coordinates associated with $q_{1,2}^{Te}$. We note that all three CDWs are incommensurate with the underlying lattice. $q_1^{CDW}$ and $q_2^{CDW}$ are related by mirror symmetry and $q_3^{CDW}$ can be connected to $q_1^{CDW}$ and $q_2^{CDW}$ by a lattice vector. These may therefore in principle correspond to a single CDW order, but for the purposes of this paper (the reason will be clear when we look at the field dependence) we treat them as independent order parameters.

The observation of a CDW in a superconductor immediately leads to the question of its relationship with superconductivity. In most instances when a CDW is found in the superconducting phase, it is an independent order parameter, which could coexist and/or compete with superconductivity (15). There is however a more interesting scenario where a CDW is a direct consequence of a periodically modulated superconducting order parameter or a pair density wave (PDW) phase (16-21). A PDW is a new phase of matter where the superconducting order parameter shows periodic spatial oscillations. A unidirectional PDW state can coexist with a uniform superconductor. In this scenario, a PDW with wave-vector **Q** is expected to generate a CDW at both **Q** and **2Q** (18). In a scenario with three CDWs, we would invoke PDWs with three primary ordering wave vectors **Q**$_i$ (i=1,2,3), which coexist with the uniform SC state. The associated CDWs should then show the same primary ordering wave vectors as well as their linear combinations ("higher harmonics") playing the role of the above mentioned **2Q** component. PDWs have been proposed to exist in superconductors with an in-plane field (22,23), but zero field PDWs require strong interactions. Experimental data showing evidence for this exotic state has been limited to cuprates (24-30) and more recently to kagome superconductors (31-35).

To investigate the relationship between the CDWs and superconductivity, we study the effect of magnetic fields on the CDW. Since the magnetic field is perpendicular to the (011) cleave plane and to the a-axis, it makes an angle of $23.7°$ with the b-axis (Fig. 3a). Extremely high magnetic fields (~40 T) oriented along this direction gives rise to the mysterious Lazarus superconducting phase or the field-polarized superconducting phase

(3). Remarkably, at our much smaller fields we see a peculiar response of the CDWs to the magnetic field. Fig. 3b-c show FFTs at 0 T and 10.5 T. The data in Fig. 3 (at 10.5 T) was obtained on the same sample with the same tip, with identical setpoint and tunnel current as the data in Fig. 2 (i.e., at 0 T). First, we find that all three CDWs are significantly suppressed in field. This can be seen by comparing the FFT at 10.5 T with the FFT at 0 T, as illustrated by linecuts obtained along the three directions (Fig. 3d-f). Equally interestingly, we find that the CDWs in field are not equally suppressed, i.e., $q_2^{CDW}$ is suppressed much more strongly than $q_1^{CDW}$, breaking mirror symmetry. The mirror symmetry breaking is not confined to just one energy, as shown by the energy dependent plot of the CDW intensity (Fig. 3g-i). This phenomenology was confirmed with a separate tip and sample combination (see supplement).

Interestingly, the suppression of the CDW with field is further enhanced when the magnetic field is tilted slightly with respect to the [011] direction. We can generate such a tilt in the sample surface while mounting the sample on the sample holder. Although these angles at present cannot be tuned controllably, they can provide valuable insights for a crystal like UTe$_2$ whose superconducting properties are highly sensitive to magnetic field orientation (3). Fig. 4 shows magnetic field dependent measurements obtained on one such fortuitous sample with a 11° tilt. Fig. 4b-d show a series of FFTs of topographies obtained at selected magnetic fields and Fig. 4g-I show FFT linecuts obtained along the three different momentum space directions. The CDW peak intensities are once again suppressed with magnetic field, eventually disappearing around 10 T. This is captured in Fig. 4e which plots the intensity of the different CDW peaks as a function of field. We find that the CDW order parameter is concomitantly suppressed with superconductivity. This phenomenology was confirmed with other tip-sample combinations. The complete disappearance of the CDW at the 11° tilted field is consistent with the lower Hc2 value when the field is applied in this direction (3).

Conventionally, the CDW order is a periodic modulation of the local charge density and as such it is not expected to couple significantly to an external magnetic field except in unconventional cases such as in the Kagome system, where there are preliminary indications of an exotic chiral CDW state (presumably carrying local orbital currents) which reverses chirality in field (32,35-37). This leaves us with two unexplained phenomena: 1) the suppression of the CDWs with magnetic field and 2) the asymmetric behavior of the two mirror-symmetry-related CDWs with field. To understand our data, we construct a Ginzburg-Landau theory which considers a triplet superconductor suggested by the symmetries of UTe$_2$. We construct a model for triplet pair density wave (PDW) orders that coexist with a uniform triplet superconductor order. In this scenario, the CDWs occur as "daughter" orders of the superconducting orders:

$$\rho_{q_i^{CDW}} \propto \vec{\Delta}_{+q_i^{CDW}} \cdot \vec{\Delta}_0^* + \vec{\Delta}_0 \cdot \vec{\Delta}^*_{-q_i^{CDW}} + h.c.,$$

where, $\vec{\Delta}_{\pm q_i^{CDW}}$ is the PDW order parameter with wavevector $\pm q_i^{CDW}$, and $\vec{\Delta}_0$ is the uniform triplet superconductivity order parameter. A Ginzburg-Landau theory analysis of these orders leads to the following conclusions: i) Above the upper critical field of the superconducting orders, the PDW and uniform superconducting orders are suppressed, as are the daughter CDWs; ii) Due to the triplet nature of the superconducting order parameter, the critical magnetic field is direction dependent; and iii) Since $\vec{\Delta}_{+q_1^{CDW}}$ and $\vec{\Delta}_{+q_2^{CDW}}$ are related by mirror symmetry, one of the two corresponding mirror related CDWs will be more suppressed than the other when mirror symmetry is broken by an external magnetic field. The responses of the CDW in magnetic field as seen by our experiments are therefore well explained by a coexisting triplet PDW order.

It is important to ask if there might be other explanations for our data. There are in fact very few alternative explanations for a CDW that is sensitive to magnetic fields. Two other possibilities are that the CDW is a daughter order of a spin density wave (SDW), or that the CDW itself has a finite angular momentum, similar to what has been observed in Kagome superconductors. While both these scenarios might explain the dependence of the CDW on a magnetic field, they each have limitations. First, no static magnetic order has been observed in in $UTe_2$ by other experimental probes (1,13,38), which makes the SDW explanation unlikely. Second, the fact that the critical field for the CDW suppression is close to critical field of the superconductor is difficult to explain with either the SDW or the finite angular momentum CDW possibility. In a nutshell, our data is most consistent with the existence of a triplet PDW state in $UTe_2$. Additional theory as well as experimental tests are important to unequivocally establish this scenario.

There are two further points to mention. Our preliminary temperature dependent measurements indicate that the CDW (as seen in the FFT) survives to 4 K and disappears somewhere between 4 K and 10 K. This is not inconsistent with the PDW scenario since it can melt to a CDW phase which can survive at higher temperatures. The other question concerns the identification of the observed peaks with the primary ordering peaks ("**1Q**") or their linear combinations **("2Q")**. We believe that the peaks we observe are the primary **1Q** peaks since we don't see any peaks at smaller q-vectors in the FFT. Consistent with this, all linear combinations of the primary peaks and Bragg peaks are also observed in our data (see supplement).

To conclude, we report the observation of incommensurate CDW orders concomitant with superconductivity in $UTe_2$. Strikingly, we observe that the CDWs are strongly affected by an external magnetic field and vanish at the $H_{c2}$ of the superconducting order. This last observation clearly implies that the CDW and superconducting order parameters are not merely coexisting but are in fact strongly coupled. This, combined with the established presence of a uniform triplet order, suggests that the superconducting state of $UTe_2$ has a triplet PDW component, which necessitates strong interactions. Finally, since STM probes the surface, the natural question is whether these orders also observed in the bulk. This calls for bulk measurements like low temperature x-ray scattering studies. If confirmed by bulk probes, our data imply that $UTe_2$ should have a complex phase diagram with triplet PDW and superconducting orders intertwined with each other. Our experiments potentially represent the first observation of a triplet PDW component of a (triplet) superconducting state which represents new phase that can occur in strongly correlated systems.

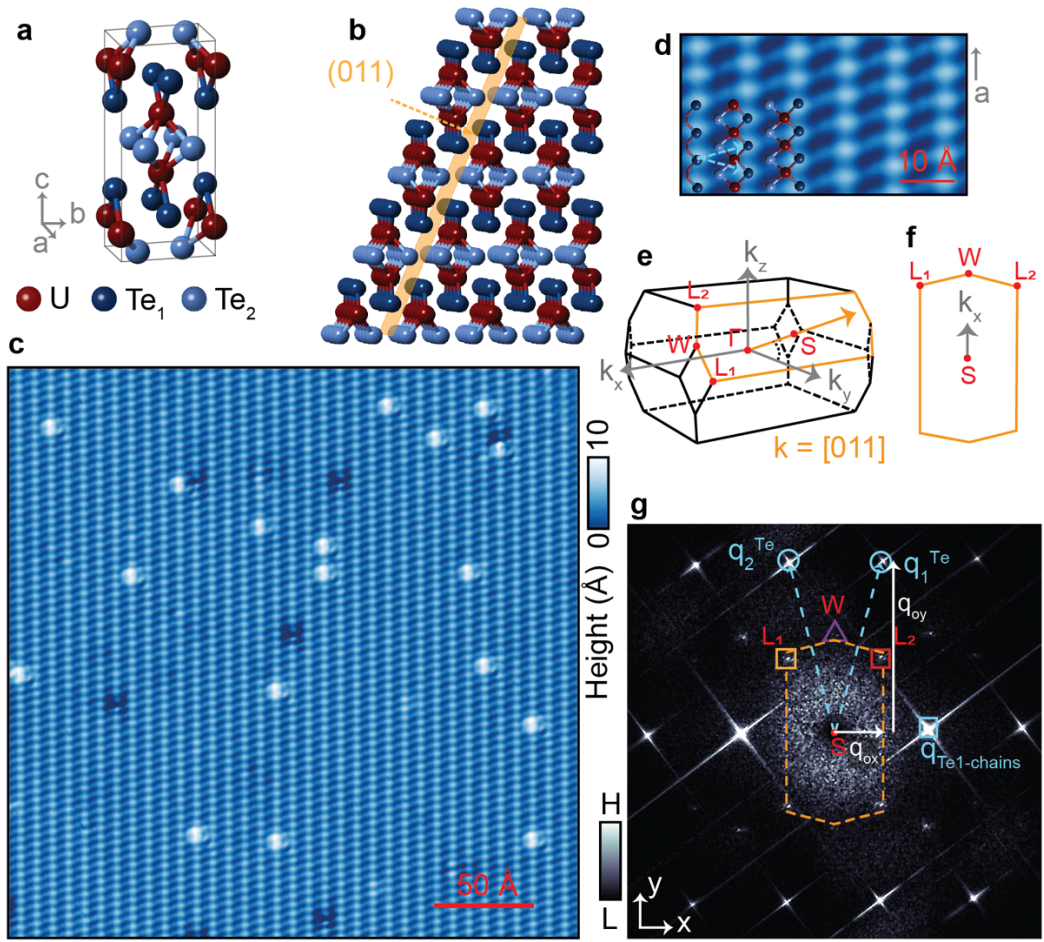

**Figure 1. Crystal structure, (011) cleave surface in real space and Fourier space**
**a,** Unit cell of $UTe_2$, consisting of U-U dimers and two inequivalent Te sites labelled as $Te_1$ and $Te_1$ depicted here by two different shades of blue. **b,** (011) easy cleave plane indicated by the yellow lattice plane. Cleaving along (011) exposes a plane of $Te_1$ and $Te_1$ atoms, with the U right underneath. **c,** A large area, atomically resolved topography obtained at T = 300 mK showing the $Te_1$ and $Te_2$ atoms (V = -60 mV, I = 200 pA) with the $Te_2$ chains being more prominent. Scale bar is 50 Å. **d,** A high-resolution, zoomed-in view of the atomic lattice (V = -60 mV, I = 200 pA). A schematic of the lattice has been overlayed on top to show the relative positions of the atoms. Scale bar is 50 Å. The cyan dashed arrows indicate the primitive lattice vectors. **e,** Schematic of the first BZ of an orthorhombic crystal with $k_x$, $k_y$ and $k_z$ directions indicated by grey arrows. The relevant points in the BZ are labelled in red. The orange hexagon is the (011) plane in reciprocal space. **f,** Schematic of the BZ with the momentum space points $L_1$, $L_2$ and W labelled in red. **g,** Fourier transform of the topography shown in **c**. The Bragg peak indicated by the cyan circle comes from the $Te_1$-$Te_1$ spacing (in the y-direction) along the chains (which is the same as the $Te_2$-$Te_2$ and and U-U distances along the chain). The Bragg peak within the cyan square comes from the inter-chain spacing (in the x-direction). The reciprocal lattice vectors are shown by the dashed arrows. The schematic of the BZ is overlayed on the FFT. The orange and red squares, and the purple triangle indicate the positions where the extra CDW peaks are observed.

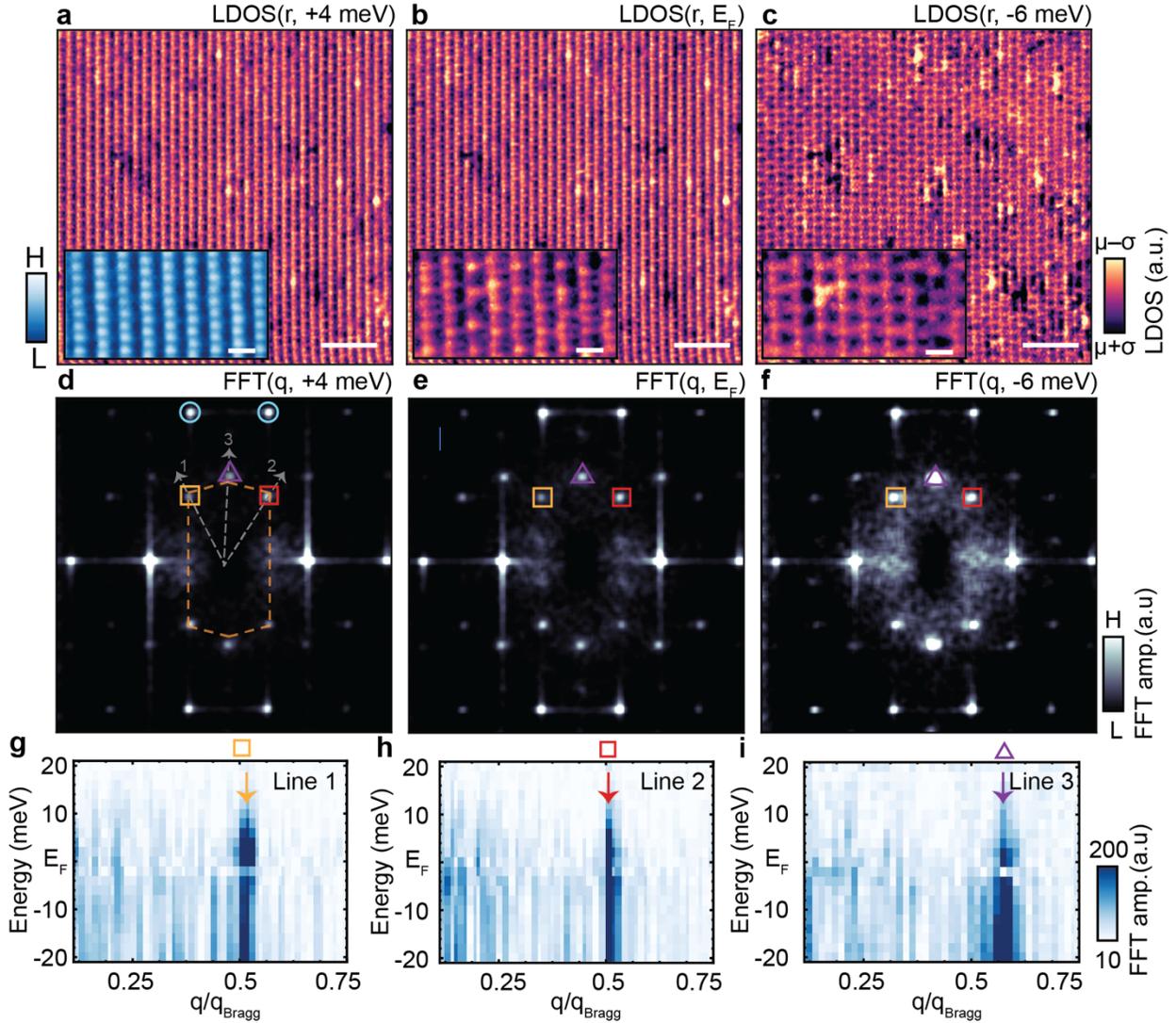

**Figure 2. Spectroscopic imaging of the three distinct charge density wave orders**
**a-c,** LDOS maps obtained on a 30 nm x 30 nm area at various energies. (Refer to the supplement for the complete set and the topography). The energies are mentioned on the top right of each map. Scale bar is 50 Å. (T = 300 mK, Tunneling setpoint: V = 50 mV, I = 250 pA). Inset of **b** and **c** show atomically resolved LDOS maps at similar energies. The corresponding topography is shown as an inset to **a**. Scale bar is 10 Å. **d-f,** Fast Fourier Transform (FFTs) of the LDOS maps shown in **a-c**. The orange dashed hexagon indicates the BZ. The cyan circles indicate the Bragg peaks from the lattice. The momentum points $L_1$, $L_2$ and W are shown by the orange, red square and purple triangles respectively. The $q_i^{CDW}$'s are very close to the BZ vertices. **g-i,** Linecuts in Fourier space plotted as an intensity map along the three directions shown by grey arrows in **d**. The positions of the peaks in the FFT i.e., the q-vectors do not change with energy, consistent with a CDW. The magnitudes of $q_1^{CDW}$ and $q_2^{CDW}$ are close to half of $q_{Te}$ (i.e. $2a_{Te-Te}$ or $2a_{U-U}$ in real space). The magnitude of $q_3^{CDW}$ is ~$0.55 q_{Te}$ (~$1.78 a_{Te-Te}$ or ~$1.78 a_{U-U}$ in real space).

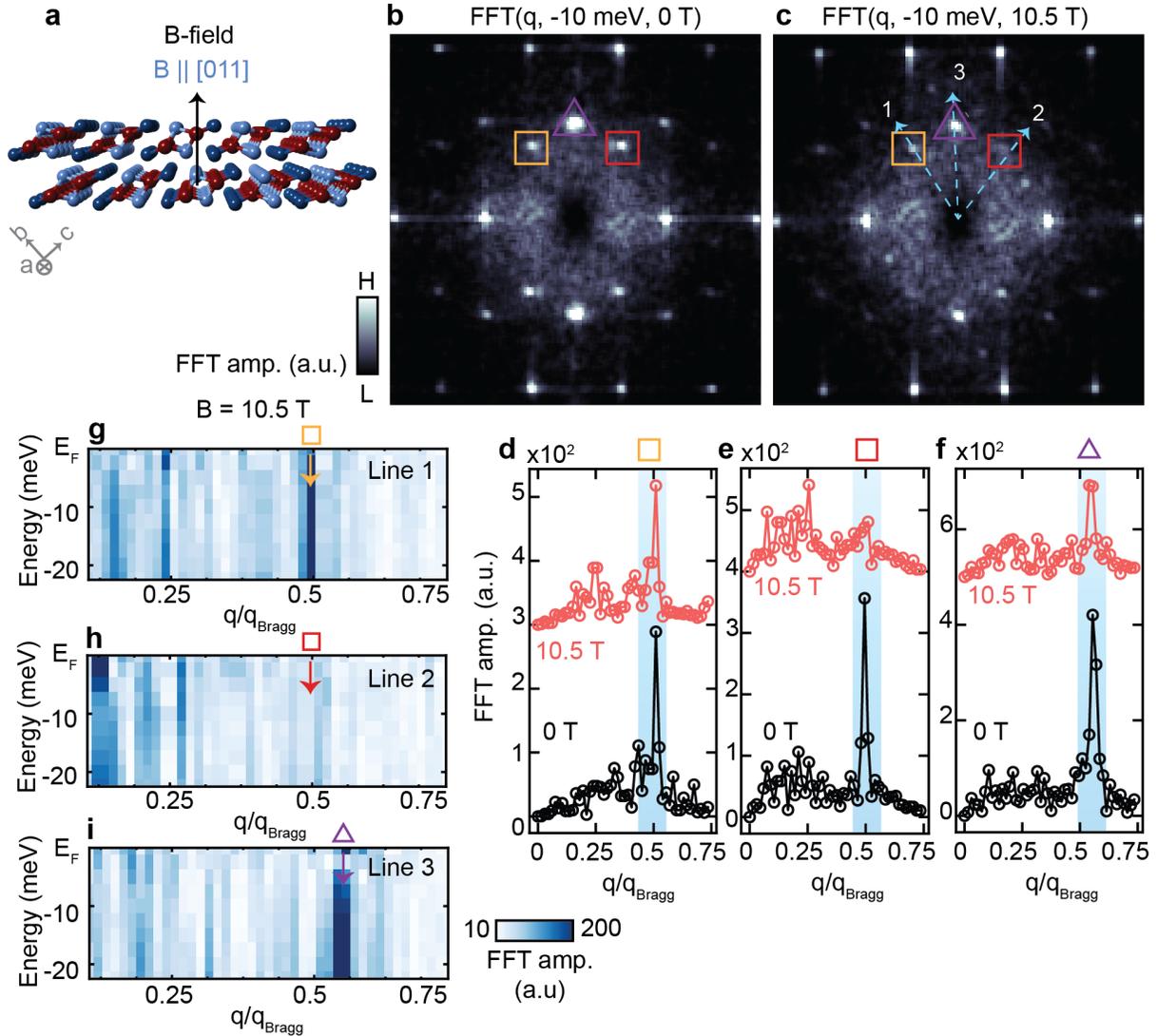

**Figure 3. Suppression and mirror symmetry breaking of the CDWs in a perpendicular magnetic field.**
**a**, Schematic showing the direction of the applied magnetic field with respect to the (011) plane. **b,c**, FFTs of LDOS maps at 10.5 T with the CDWs marked. The data were obtained on the same area as the 0 T data shown in figure 2 and obtained at -10 meV (Tunneling setpoint: V = 50 mV, I = 250 pA). The color scale has been kept identical for **b** and **c**. **d-f,** Linecuts of the Fourier transforms of the LDOS maps at a single energy (-10 meV), along the three different directions of the CDWs, $q_{L_1}$, $q_{L_2}$ and of $q_W$ (indicated by the cyan arrows in **c**) for the FFT at 0 T and 10.5 T, respectively. While all the CDW peaks are significantly suppressed, **e** ($q_{L_2}$) shows a much stronger suppression in comparison to **d**. This reveals a putative breaking of the mirror symmetry in the presence of a magnetic field. **g-i,** Linecuts in of FFTs at 10.5 T at different energies along the three directions indicated by the cyan arrows in **c**, plotted as an intensity map. One can visually see that the Fourier amplitude at $q_{L_2}$ in **f** is highly suppressed compared to the 0 T data in Fig.2h.

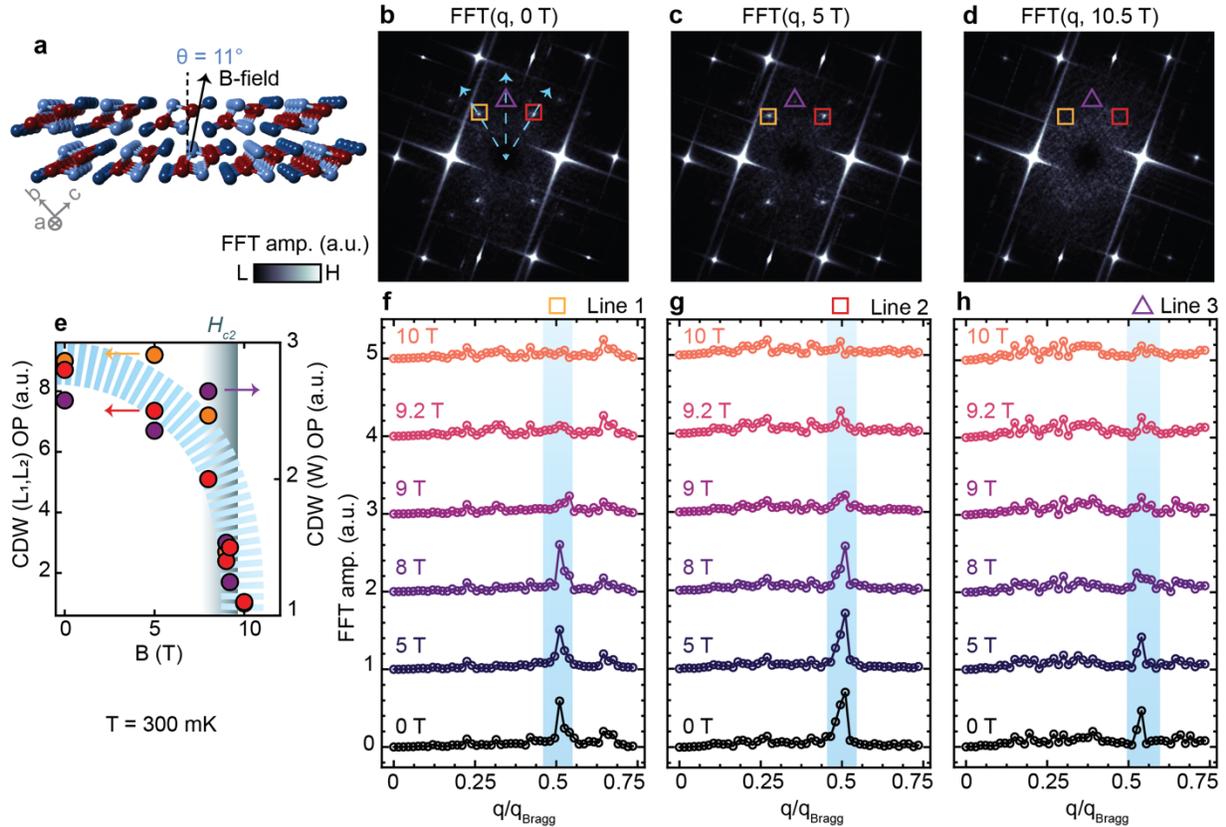

**Figure 4. Disappearance of the CDWs above $H_{c2}$ for magnetic field tilted at 11 degrees to the [011] direction.**
**a**, Schematic of the direction of the applied magnetic field with respect to the (011) plane. The magnetic field is tilted 11° with respect to the [011] direction. **b-d**, FFTs obtained at 0 T, 5 T and 10 T respectively, on the same area with identical settings at T = 300 mK. (V = -40 mV, I = 120 pA). The CDWs are marked by the orange and red squares and purple triangle. The color scale has kept constant for **b-d**. **e**, Plot of the strength of the CDW order parameter (Fourier amplitude) as a function of magnetic field. The markers are color-coded to represent the respective CDWs. Blue dashed region is a guide to the eye. The $H_{c2}$ is indicated by the grey region. **f-h**, Fourier transform linecuts obtained along 3 different directions of the CDWs as a function of magnetic field, showing clear suppression of the peak amplitudes above 9 T.

## Methods

Single crystals of $UTe_2$ were used for this. The growth and characterization are mentioned in detail elsewhere (1). The crystal orientation was determined by Laue diffraction. Samples were cleaved in situ at ~90 K and in an ultrahigh-vacuum chamber. After cleaving, the samples were directly transferred to the STM head. STM measurements were performed using a Unisoku STM at an instrument temperature of 300 mK (unless otherwise specified) using chemically etched and annealed tungsten tips. The temperature values reported were measured at the $^3$He pot; the actual sample temperature could be slightly higher. d$I$/d$V$ spectra were collected using a standard lock-in technique at a frequency of 913 Hz.

## Acknowledgements

The authors thank Steven Kivelson and Qimiao Si for useful discussions. STM studies at the University of Illinois, Urbana-Champaign was supported by the U.S. Department of Energy (DOE), Office of Science, Office of Basic Energy Sciences (BES), Materials Sciences and Engineering Division under Award No. DE-SC0022101. V.M. acknowledges partial support from Gordon and Betty More Foundation's EPiQS Initiative through grant GBMF4860 and the Quantum Materials Program at CIFAR where she is a Fellow. Theoretical work was supported in part by the US National Science Foundation through the grant DMR 1725401 at the University of Illinois (E.F., L.N.) and by a fellowship of the Institute for Condensed Matter Theory of the University of Illinois (L.N.). J.M.M. thanks ARO MURI Grant No. W911NF2020166 for support. Research at the University of Maryland was supported by the Department of Energy Award No. DE-SC-0019154 (sample characterization), the Gordon and Betty Moore Foundation's EPiQS Initiative through Grant No. GBMF9071 (materials synthesis), the National Science Foundation under Grant No. DMR-2105191 (sample preparation), the Maryland Quantum Materials Center, and the National Institute of Standards and Technology.

## Author contributions

A.A. and V.M. conceived the experiments. The single crystals were provided by S.R., S.R.S., J.P. and N.P.B. M.R. performed the Laue characterization of the single crystals. A.A. and A.R. obtained the STM data. A.A. and V.M. performed the analysis and J.M.M., L.N. and E.F. provided the theoretical input on the interpretation of the data. A.A., V.M., J.M.M. and E.F. wrote the paper with input from all authors.

## Competing Interests

The authors declare no competing interests.

# Supplementary information for
# Magnetic-field sensitive charge density wave orders in the superconducting phase of UTe$_2$

Anuva Aishwarya[1], Julian May-Mann[1], Arjun Raghavan[1], Laimei Nie[1], Marisa Romanelli[1], Sheng Ran[2,3], Shanta R. Saha[2], Johnpierre Paglione[2], Nicholas P. Butch[2,3], Eduardo Fradkin[1], Vidya Madhavan[1]

[1]Department of Physics and Materials Research Laboratory, University of Illinois at Urbana-Champaign, Urbana, IL, USA.
[2]Center for Nanophysics and Advanced Materials, Department of Physics, University of Maryland, College Park, MD, USA
[3]NIST Center for Neutron Research, National Institute of Standards and Technology, Gaithersburg, MD, USA.

Supplementary Figures 1- 9
Ginzburg Landau description of the triplet pair density wave
References

**S1. Supplementary Figures**

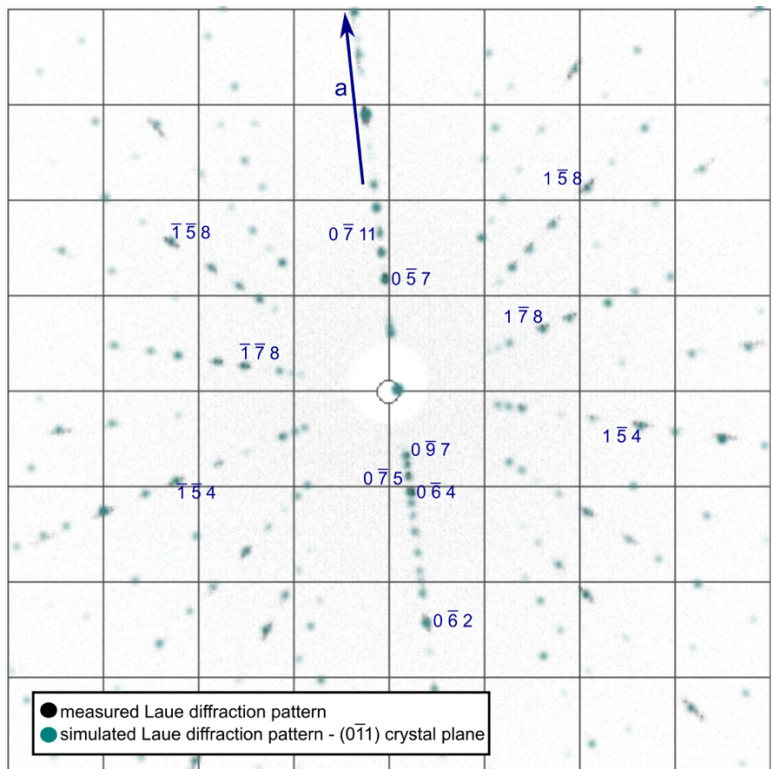

**Supplementary Figure 1. Laue diffraction from the (011) aligned crystal**

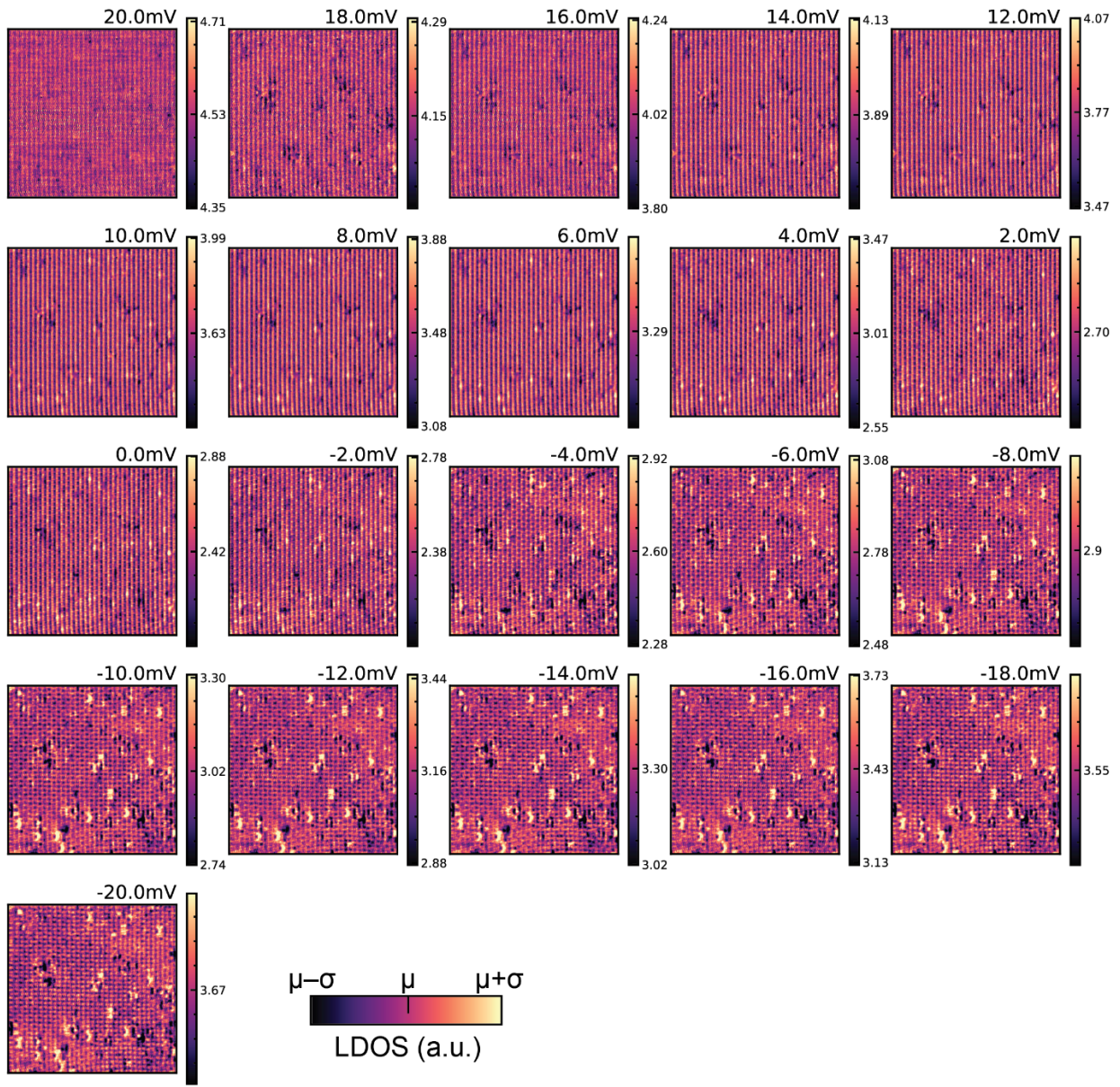

**Supplementary Figure 2. LDOS at 300 mK**
LDOS maps obtained at several energies above and below $E_F$.

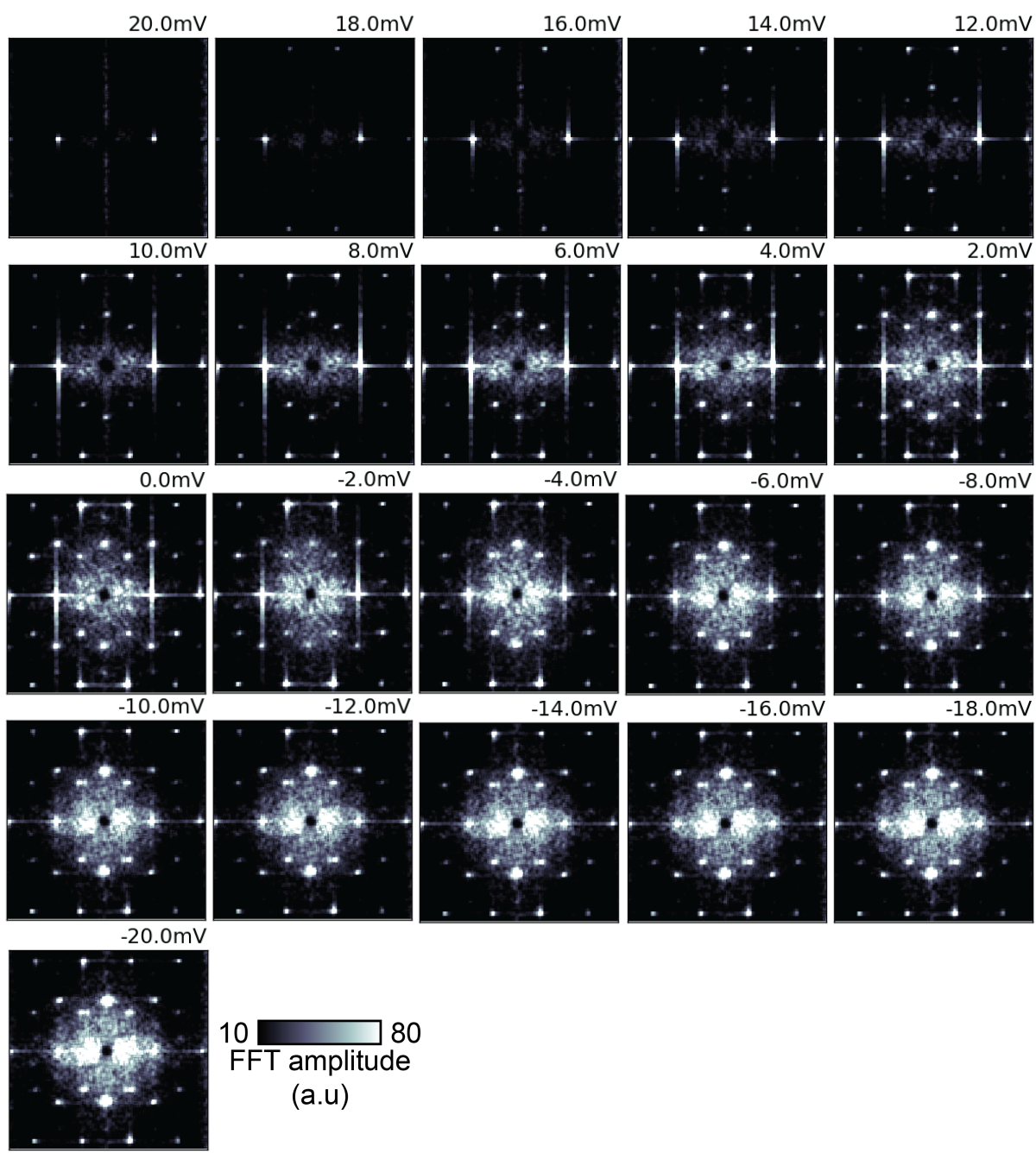

**Supplementary Figure 3. FFT of LDOS at 300 mK**
FFTs of LDOS maps obtained at several energies above and below $E_F$.

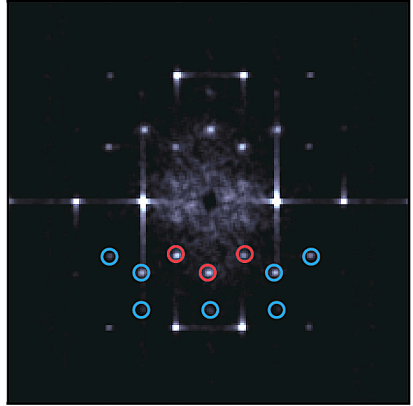

**Supplementary Figure 4. FFT showing primary and secondary CDW peaks.**
FFT at the $E_F$ where the primary and secondary CDWs are shown using red circles and blue circles respectively.

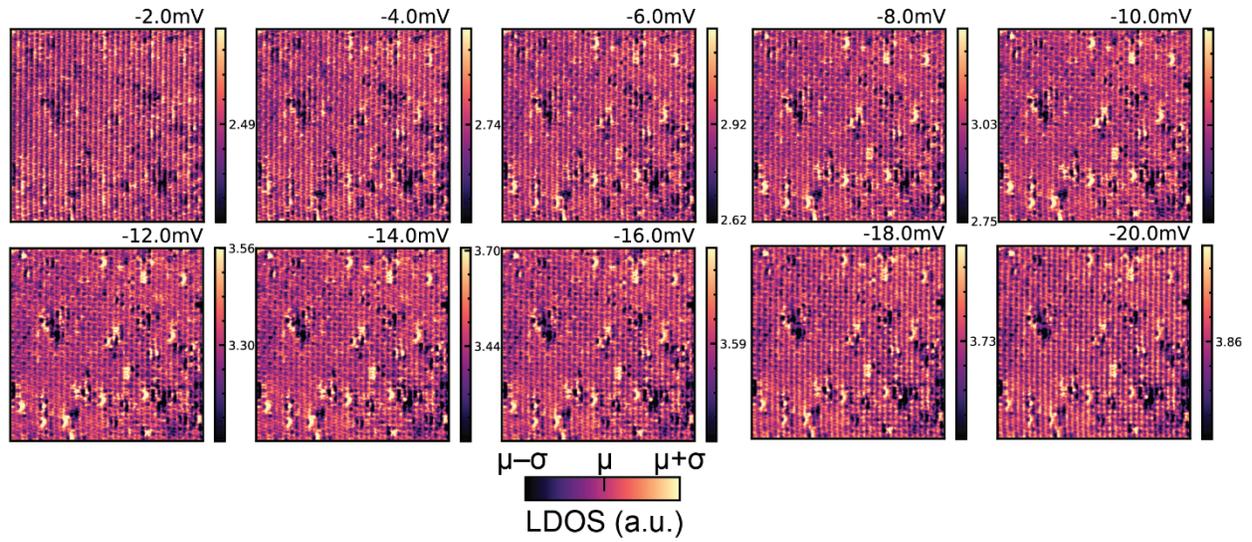

**Supplementary Figure 5. LDOS in presence of a 10.5 T magnetic field**
LDOS maps obtained at several energies in a perpendicular magnetic field.

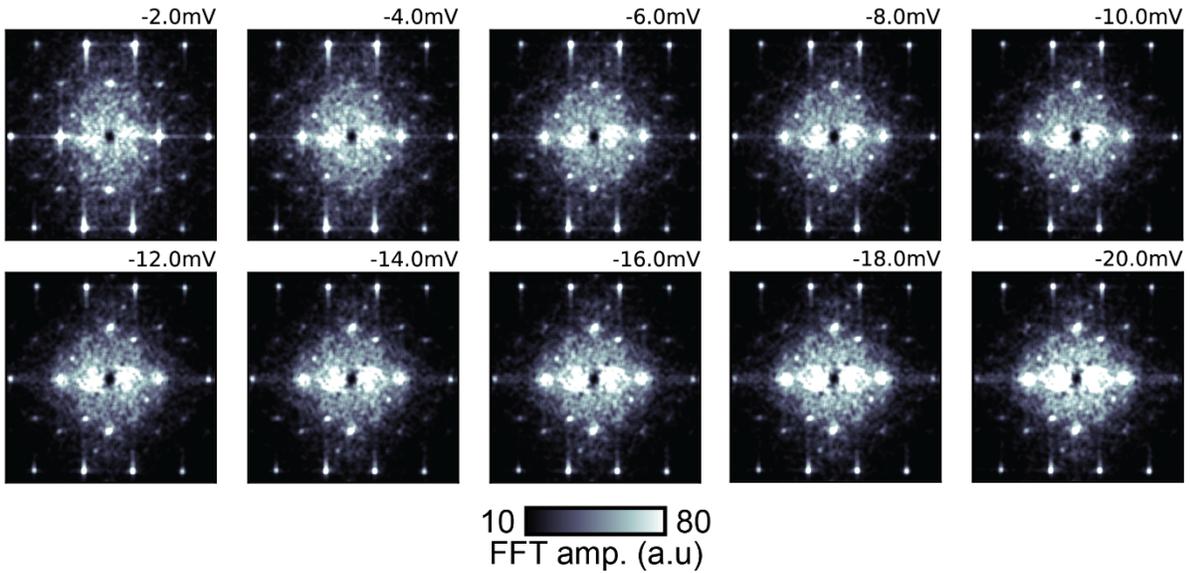

**Supplementary Figure 6. FFT of LDOS in presence of a 10.5 T magnetic field**
FFTs of LDOS maps obtained at several energies in a perpendicular magnetic field.

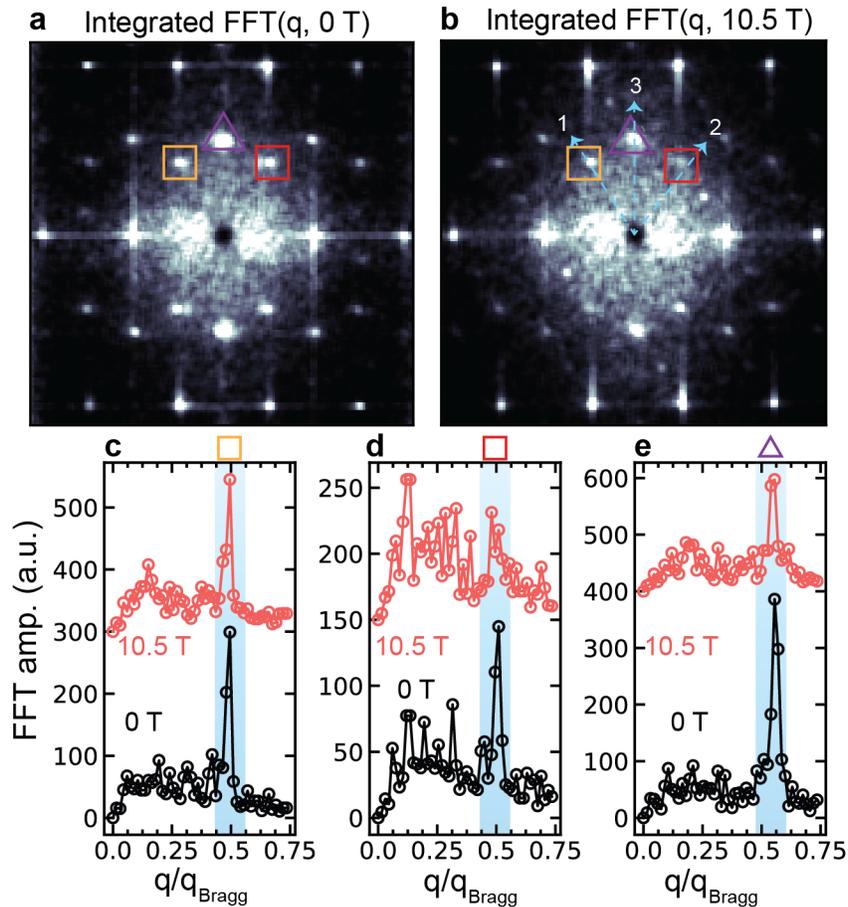

**Supplementary Figure 7. Partial suppression and mirror symmetry breaking of the CDWs in the integrated FFT signal.**
**a-b**, Comparison of FFTs of integrated signal obtained from integrating LDOS maps below $E_F$ for a 0 T field and 10.5 T field. The FFT of the integrated signal also shows similar behavior as the FFT of individual energy slices. **c-e**, Linecuts obtained along 3 different directions for the 3 CDWs illustrating the mirror symmetry breaking. **d** is clearly more suppressed than **c**.

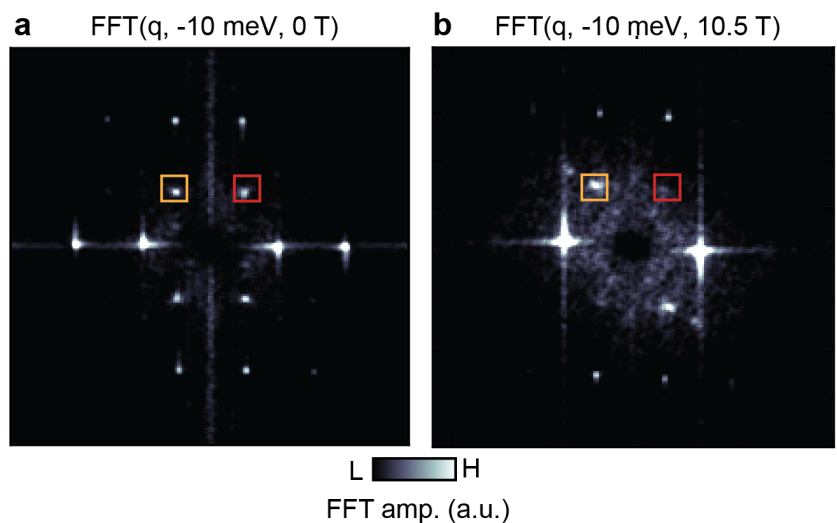

**Supplementary Figure 8. Reproducibility of the partial suppression of the CDW in a 10.5 T perpendicular magnetic field across samples.**

**a-b**, Additional dataset obtained with a different tip-sample combination showing the partial suppression and mirror symmetry breaking of the CDW in a perpendicular magnetic field. The FFTs shown have the same intensity scale.

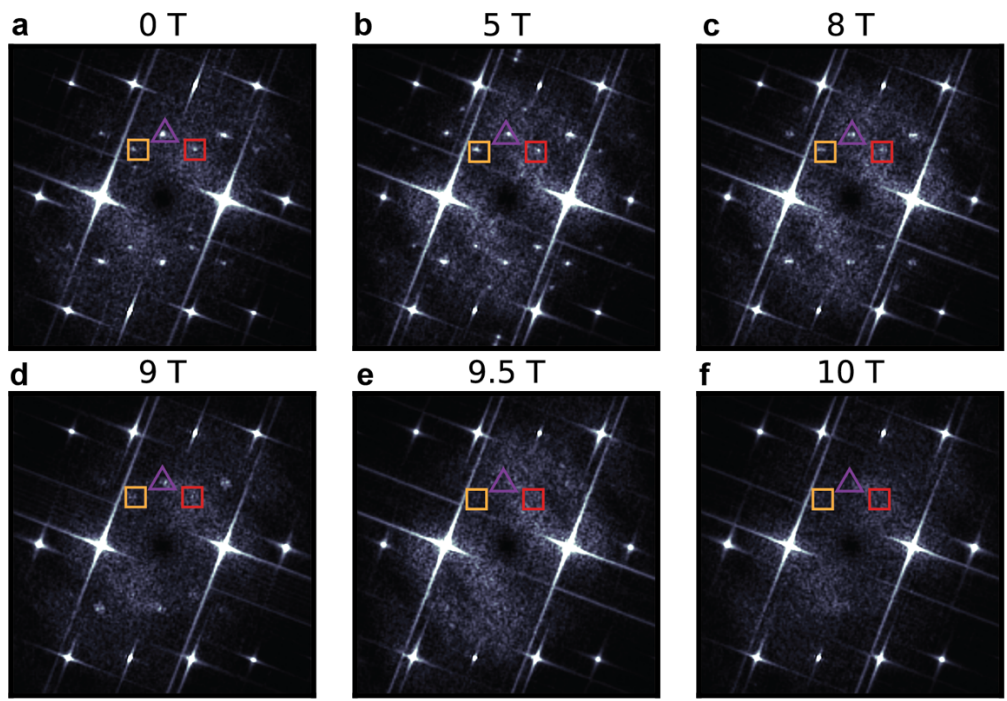

**Supplementary Figure 9. Reproducibility of the suppression of the CDWs in a 11-degree tilted magnetic field with a different tip.**

**a-f,** Additional dataset showing a series of FFT of topographies obtained as a function of increasing magnetic field at 11 degrees with respect to the [011] direction with a different tip. (V = 50 mV, I = 150 pA). The intensity scale of all FFTs has been kept constant. The surface tilt is measured using the tilt correction function in the Nanonis module.

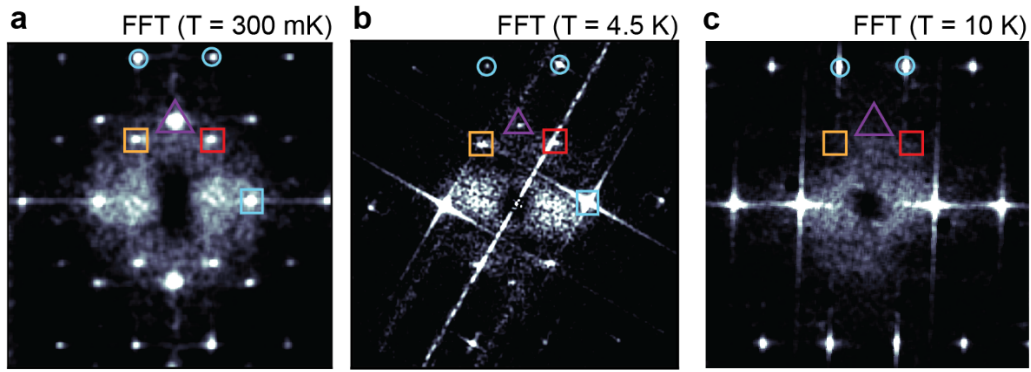

**Supplementary Figure 9. Melting of the CDWs as a function of temperature**
**a-c**, FFTs of LDOS maps obtained as a function of temperature. The CDWs persist till 4 K and are gone by 10 K.

## S2. Ginzburg-Landau description of the triplet pair density waves

In this supplement we consider a Ginzburg-Landau description of triplet pair density waves (PDWs) coexisting with a uniform triplet superconductor in a material with the same symmetries as UTe$_2$. As we shall show, this theory predicts the existence of "daughter" charge density waves (CDWs) that are suppressed in an external magnetic field.

## The Triplet Pair Density Wave

Before considering the full Ginzburg-Landau theory, it will be useful to give an overview of the triplet PDW state. In real space, we expand the local triplet Cooper pair amplitude as

$$\langle c_\sigma(r) c_{\sigma'}(r') \rangle = i \, [\tau_y \vec{\tau}]_{\sigma\sigma'} \cdot [\vec{\Delta}_0(r-r') + \sum_Q \vec{\Delta}_Q(r-r') \, e^{iQ \cdot (r+r')/2} ], \quad (1)$$

where $\tau_i$ (with $i = x, y, z$) are the three 2 × 2 Pauli matrices, $r$ and $r'$ label the coordinates of the two electrons that form the Cooper-pair, and the possible ordering wavevectors (and their harmonics) are given by Q. Here, $\vec{\Delta}_0$ is the uniform triplet superconductor, while $\vec{\Delta}_Q$ is the triplet PDW with wavevector $Q$. Due to the fermion anti-commutation relationships, $\vec{\Delta}_0$ and $\vec{\Delta}_Q$ must be odd functions of $r - r'$. In momentum space, the above equation becomes

$$\langle c_\sigma(q/2+k) c_{\sigma'}(q/2-k) \rangle = i \, [\tau_y \vec{\tau}]_{\sigma\sigma'} \cdot [\vec{\Delta}_0(k) \delta^3(q) + \sum_Q \vec{\Delta}_Q(k) \, \delta^3(q-Q)] \quad (2)$$

where $k$ is the relative momentum of the two electrons, $q$ is the total momentum of the Cooper-pairs, and $\delta$ is the Dirac delta function. The vectors, $\vec{\Delta}_0$ and $\vec{\Delta}_Q$, are both odd functions of $k$. In the limit where $Q \to 0$, $\vec{\Delta}_Q$, is equivalent to the uniform triplet superconductor $\vec{\Delta}_0$. The PDW state has the same periodic modulation of a Larkin-Ovchinnikov SC state (1) but in the absence of an external magnetic field, i.e., without explicit breaking of time reversal invariance, and the period of the PDW is thus not tuned by an external magnetic field. (For a review see (2))

In a real material, $\vec{\Delta}_0$ and $\vec{\Delta}_Q$, should form parity-odd irreducible representations (irreps) of the crystal space group. For UTe$_2$, the space group is $D_{2h}$, which is characterized by three mirror symmetries, $M_x$, $M_y$, and $M_z$ (it is also possible to equivalently characterize $D_{2h}$ in terms of three $C_2$ rotations). There are 4 parity-odd irreps of $D_{2h}$ (3), which are referred to as $A_u$, $B_{1u}$, $B_{2u}$, and $B_{3u}$. Their transformation properties of the triplet PDW, $\vec{\Delta}_Q$, under $M_x$ are

$$M_x: \vec{\Delta}_Q \to \vec{\Delta}_{M_xQ} \text{ for } \vec{\Delta}_Q \in B_{1u}, B_{2u},$$

$$M_x: \vec{\Delta}_Q \to -\vec{\Delta}_{M_xQ} \text{ for } \vec{\Delta}_Q \in A_u, B_{3u},$$

(3)

where $M_xQ$ is the $M_x$ mirror transformed wavevector $Q$. The transformation properties under $M_y$ are

$$M_y: \vec{\Delta}_Q \to \vec{\Delta}_{M_yQ} \text{ for } \vec{\Delta}_Q \in B_{1u}, B_{3u},$$

$$M_y: \vec{\Delta}_Q \to -\vec{\Delta}_{M_yQ} \text{ for } \vec{\Delta}_Q \in A_u, B_{2u}.$$

(4)

The transformation properties under $M_z$, are

$$M_z: \vec{\Delta}_Q \to \vec{\Delta}_{M_zQ} \text{ for } \vec{\Delta}_Q \in B_{2u}, B_{3u},$$

$$M_z: \vec{\Delta}_Q \to -\vec{\Delta}_{M_zQ} \text{ for } \vec{\Delta}_Q \in A_u, B_{1u}.$$

(5)

The transformation properties of the uniform component, $\vec{\Delta}_0$, are related to those above by taking $Q = M_xQ = M_yQ = M_zQ = 0$.

### Ginzburg-Landau Theory

In this section, we consider a Ginzburg-Landau theory of three triplet PDWs coexisting with uniform triplet superconductivity in a material with $D_{2h}$ symmetry (the same symmetry as UTe$_2$). As we shall show, this theory leads to daughter charge density wave (CDWs) that are suppressed in a magnetic field. Furthermore, when the triplet superconducting orders have finite angular momentum, the suppression is anisotropic, and uneven for different CDWs.

The Ginzburg-Landau theory is constructed from a uniform triplet order parameter $\vec{\Delta}_0$ and triplet PDW order parameters $\vec{\Delta}_{\pm Q_i}$ with wavevector $\pm Q_i$, and $i = 1,2,3$. The wavevectors $Q_i$ are defined such that for the cleave surface of the material (defined the same way as in the main text) the wavevectors $Q_i$ project onto $q_i^{CDW}$. The PDW order parameters therefore project onto surface PDW order parameters of the form $\vec{\Delta}_{\pm Q_i} \to \vec{\Delta}_{\pm q_i^{CDW}}$ on the cleave surface. Since $M_x$ mirror symmetry is the only symmetry preserved by the cleave surface, we will primarily consider the transformation properties of the order parameters under $M_x$. On the cleave surface, mirror symmetry acts as $M_x: q_1^{CDW} \to -q_2^{CDW}$ and $M_x: q_3^{CDW} \to -q_3^{CDW}$, and so we take $Q_i$ to be defined such that $M_x: Q_1 \to -Q_2$ and $M_x: Q_3 \to -Q_3$, as well.

The Landau free energy density for $\vec{\Delta}_0$ and $\vec{\Delta}_{\pm Q_i}$ is given by

$$\mathcal{F} = \mathcal{F}_2 + \mathcal{F}_4$$

$$\mathcal{F}_2 = m_0 |\vec{\Delta}_0|^2 + \sum_i m_i \left(|\vec{\Delta}_{Q_i}|^2 + |\vec{\Delta}_{-Q_i}|^2\right)$$

$$\mathcal{F}_4 = \lambda_{00} |\vec{\Delta}_0|^4 + \sum_i \lambda_{0i} \left(|\vec{\Delta}_{Q_i}|^2 |\vec{\Delta}_0|^2 + |\vec{\Delta}_{-Q_i}|^2 |\vec{\Delta}_0|^2\right) \quad (6)$$

$$+ \sum_{ij} \lambda_{ij} \left(|\vec{\Delta}_{Q_i}|^2 |\vec{\Delta}_{Q_j}|^2 + |\vec{\Delta}_{-Q_i}|^2 |\vec{\Delta}_{-Q_j}|^2\right)$$

$$+ \sum_{ij} \lambda'_{ij} \left(|\vec{\Delta}_{Q_i}|^2 |\vec{\Delta}_{-Q_j}|^2 + |\vec{\Delta}_{-Q_i}|^2 |\vec{\Delta}_{Q_j}|^2\right)$$

Here, $\lambda_{ij} = \lambda_{ji}$, $\lambda'_{ij} = \lambda'_{ji}$. Due to mirror symmetry, $m_1 = m_2$, $\lambda_{01} = \lambda_{02}$, $\lambda_{i1} = \lambda_{i2}$ and $\lambda'_{i1} = \lambda'_{i2}$. For stability, $\lambda_{00}, \lambda_{ii} > 0$. To favor coexistence of the superconducting order parameters, $\lambda_{ij} < 0$ for $i \neq j$ and $\lambda'_{ij} < 0$ for all $i$ and $j$.

In the ordered phase, where $\vec{\Delta}_0$ and $\vec{\Delta}_{\pm Q_i}$ all have expectation values ($m_0, m_i < 0$), there will be daughter CDW orders,

$$\rho_{Q_i} \propto \vec{\Delta}_{Q_i} \cdot \vec{\Delta}_0^* + \vec{\Delta}_0 \cdot \vec{\Delta}_{-Q_i}^*,$$
$$\rho_{Q_i+Q_j} \propto \vec{\Delta}_{Q_i} \cdot \vec{\Delta}_{-Q_j}^* + a_i \rho_{Q_i} \rho_{Q_j}, \quad (7)$$
$$\rho_{Q_i-Q_j} \propto \vec{\Delta}_{Q_i} \cdot \vec{\Delta}_{Q_j}^* + b_i \rho_{Q_i} \rho_{-Q_j},$$

where $a_i$ and $b_i$ are complex constants, and the CDW order parameters $\rho_Q$ are a complex scalar field that satisfies $\rho_Q^* = \rho_{-Q}$. On the cleave surface, the CDW operators project onto surface CDW operators $\rho_{Q_i} \to \rho_{q_i^{CDW}}$ and $\rho_{Q_i \pm Q_j} \to \rho_{q_i^{CDW} \pm q_j^{CDW}}$.

In this Landau theory the daughter orders arise from the cubic couplings

$$\mathcal{F}_{CDW} = \sum_i g_{0i} \rho_{Q_i} [\vec{\Delta}_{Q_i} \cdot \vec{\Delta}_0^* + \vec{\Delta}_0 \cdot \vec{\Delta}_{-Q_i}^*]$$

$$+ \sum_{ij} g_{ij} \rho_{Q_i+Q_j} \vec{\Delta}_{Q_i} \cdot \vec{\Delta}_{-Q_j}^* + g'_{ij} \rho_{Q_i-Q_j} \vec{\Delta}_{Q_i} \cdot \vec{\Delta}_{Q_j}^* \quad (8)$$

$$+ \gamma_{ij} \rho_{Q_i+Q_j}^* \rho_{Q_i} \rho_{Q_j} + \gamma'_{ij} \rho_{Q_i-Q_j}^* \rho_{Q_i} \rho_{-Q_j} + h.c.,$$

Based on this, when $\vec{\Delta}_{Q_i}$ and $\vec{\Delta}_0$ are non-zero, the free energy is minimized when the CDWs are also non-zero. For brevity we have omitted the any quartic terms from $\mathcal{F}_{CDW}$, although such terms are allowed by symmetry. We should note that if one considers the case of a system with CDW and uniform SC coexistent orders, then a modulated ("PDW") component would be induced by the first of the cubic terms. Thus, the Landau theory does not distinguish these two scenarios.

We now add an external magnetic field $\vec{H}$ to the Landau theory. If we include a gradient term in the free energy expansion, the magnetic field minimally couples to the superconducting orders as $\left|D_\mu \vec{\Delta}_0\right|^2$, and $\left|D_\mu \vec{\Delta}_{\pm Q_i}\right|^2$, where $D_\mu = \partial_\mu - i2eA_\mu$, and $A_\mu$ is the electromagnetic gauge field. Since the superconducting orders are spin triplet, they can have a finite angular momentum that also couples to the magnetic field. The angular momenta of the superconducting orders are $i\vec{\Delta}_0 \times \vec{\Delta}_0^*$, and $i\vec{\Delta}_{\pm Q_i} \times \vec{\Delta}_{\pm Q_i}^*$, and they couple to the external magnetic field via

$$\mathcal{F}_{Mag} = -\epsilon_0\, \vec{H} \cdot \left(i\vec{\Delta}_0 \times \vec{\Delta}_0^*\right) - \sum_j \epsilon_j\, \vec{H} \cdot \left(i\vec{\Delta}_{Q_j} \times \vec{\Delta}_{Q_j}^*\right)$$
$$- \sum_j \epsilon_j\, \vec{H} \cdot \left(i\vec{\Delta}_{-Q_j} \times \vec{\Delta}_{-Q_j}^*\right). \tag{9}$$

Here, $\epsilon_0, \epsilon_j > 0$ and mirror symmetry requires that $\epsilon_1 = \epsilon_2$. When the superconducting order parameters each form a single irrep of $D_{2h}$, the angular momenta all vanish (4,5). However, if time-reversal symmetry is broken and the order parameters are combinations of different irreps of $D_{2h}$, the angular momentum can be non-vanishing, $\langle i\vec{\Delta}_0 \times \vec{\Delta}_0^*\rangle \neq 0$, $\langle i\vec{\Delta}_{\pm Q_i} \times \vec{\Delta}_{\pm Q_i}^*\rangle \neq 0$. Since $\epsilon_0, \epsilon_j > 0$, it is energetically favorable for the angular momentum to aligned with the magnetic field. In principle, when the superconducting order parameters are combinations of different irreps each irrep should correspond to a distinct term in the Ginzburg-Landau theory. Nevertheless, for simplicity, we have assumed that the Ginzburg-Landau theory can be written in terms of the superconducting orders $\vec{\Delta}_{Q_i}$ and $\vec{\Delta}_0$, instead of the distinct irreps. This simplification does not reflect the fact that different irreps will, in general, transition at different temperature. but this additional feature does not change the qualitative features of our analysis.

The superconducting orders are suppressed in an external magnetic field, and above an upper critical field the superconducting orders, and the daughter CDW orders will vanish. The Ginzburg-Landau theory therefore correctly predicts the suppression of the CDW orders in an external magnetic field. Due to the coupling between the angular momentum and magnetic field, the upper critical field will be higher when the magnetic field is aligned with the angular momentum of a superconducting order and lower when they are anti-aligned.

If $M_x$ mirror symmetry is preserved by the $\vec{\Delta}_{\pm Q_1}$ and $\vec{\Delta}_{\pm Q_2}$ PDWs, their respective angular momenta are related via

$$\hat{y} \cdot \langle i\vec{\Delta}_{\pm Q_1} \times \vec{\Delta}_{\pm Q_1}^*\rangle = -\hat{y} \cdot \langle i\vec{\Delta}_{\mp Q_2} \times \vec{\Delta}_{\mp Q_2}^*\rangle$$
$$\hat{z} \cdot \langle i\vec{\Delta}_{\pm Q_1} \times \vec{\Delta}_{\pm Q_1}^*\rangle = -\hat{z} \cdot \langle i\vec{\Delta}_{\mp Q_2} \times \vec{\Delta}_{\mp Q_2}^*\rangle \tag{10}$$

Because of this, when $H_y \neq 0$, or $H_z \neq 0$, the $\vec{\Delta}_{\pm Q_1}$ PDW will be more suppressed than the $\vec{\Delta}_{\pm Q_2}$ PDW or vice versa. Which of the two is more suppressed depends on the sign and magnitude of $H_y$ and $H_z$. This agrees with the observed mirror symmetry breaking.

We note that the Ginzburg-Landau theory predicts existence of both $\rho_{Q_i}$ and $\rho_{Q_i \pm Q_j}$ daughter CDWs. Both types of CDWs are observed in UTe$_2$, but the $\rho_{Q_i \pm Q_j}$ CDWs are weaker than the $\rho_{Q_i}$ CDWs. A possible explanation for this is that the amplitudes of the PDW order parameters are weaker than that of the uniform superconducting order parameter,

$$|\langle \vec{\Delta}_{\pm Q_i} \rangle| \ll |\langle \vec{\Delta}_0 \rangle|. \tag{11}$$

In this case the $\rho_{Q_i}$ CDWs would be the dominate charge orders, since $\rho_{Q_i}$ is linear in the PDW order parameters, while $\rho_{Q_i \pm Q_j}$ is quadratic in PDWs.